\documentclass[prc,aps,preprint,showpacs]{revtex4}
\usepackage{graphicx}

\newcommand{\simleq}{\; \raisebox{-0.4ex}{\tiny$\stackrel
{{\textstyle<}}{\sim}$}\;}
\begin{document}

\title{Enhanced $E1$ transition between weakly-bound excited states 
in the nucleus $^{27}$Ne} 

\author{ Ikuko Hamamoto$^{1,2}$ }

\affiliation{
$^{1}$ {\it Riken Nishina Center, Wako, Saitama 351-0198, Japan } \\ 
$^{2}$ {\it Division of Mathematical Physics, Lund Institute of Technology 
at the University of Lund, Lund, Sweden} }   




\begin{abstract}
Inspired by the recently-reported strong electric-dipole ($E1$) transition 
between the weakly-bound first and second excited states,  
$3/2^-$ at 765 keV and 
$1/2^+$ at 885 keV, 
in the nucleus $^{27}_{10}$Ne$_{17}$, 
the $E1$ transition is estimated in a model 
by properly taking into account the effect of 
both deformation and weakly-bound neutrons.   In addition to 
both the spin-parities, 1/2$^{+}$ and 3/2$^{-}$, and observed nearly 
degenerate energies of 
the two excited states, the observed order of magnitude of the $E1$ transition 
strength between the two states is very naturally explained  
in the case that these two excited states are prolately deformed, 
in terms of the transitions    
between the halo components of the wave functions of 
the weakly-bound odd-neutrons, $s_{1/2} \rightarrow p_{3/2}$ 
and $s_{1/2} \rightarrow p_{1/2}$,   
in addition to  
the large probability of the p$_{3/2}$ component 
in the weakly-bound neutron [330 1/2] 
orbit. 
The large probability is the result of the shell-structure unique 
in weakly-bound 
or resonant neutrons.

\end{abstract}


\maketitle

\newpage

\section{INTRODUCTION} 
The study of nuclei far from the $\beta$ stability line has shown many
new interesting phenomena which are unexpected from our common sense about
stable nuclei; among others, halo phenomena, the change of the shell structure, 
$\ell$-$s$ splittings and magic numbers. 
In the present work I study the unusually strong $E1$ transition between 
the two excited states of the nucleus $^{27}_{10}$Ne$_{17}$, 
by properly taking into account the possible deformation 
and the weakly-bound odd neutron. 

Though low-energy $E1$ transitions are often observed in $\gamma$ decays of 
nuclei, in both spherical and deformed nuclei 
their strength is usually orders of magnitude  
weaker than the Weisskopf unit.   This is due to several reasons: 
(a) The isoscalar part of the dipole operator corresponds 
to the center of mass motion and thus does not contribute to $E1$ transitions; 
(b) Since the isovector part of the residual dipole-dipole interaction is 
repulsive, the major part of the unperturbed $E1$ strength lying originally 
in the low energy region is shifted to the region of higher energy, 
especially to the giant dipole resonance (GDR);    
(c) There is no appreciable amount of low-energy $E1$ strength 
in one-particle excitations of spherical stable nuclei 
due to the nuclear shell structure.  
If the nuclear potential is expressed by a harmonic oscillator potential, 
$E1$ transitions which connect the states with different parities do not occur 
between the one-particle orbits belonging to the same major shell. 
In realistic nuclear potentials the parities of one-particle orbits 
in a given major shell of lighter nuclei with 2$<$N$\leq$8, 8$<$N$\leq$20 and  
20$<$N$\leq$28, 
are the same, while in the major shells of heavier nuclei 
28$<$N$\leq$50, 
50$<$N$\leq$82, and 82$<$N$\leq$126 ...,  
so-called high-j orbits such as 1g$_{9/2}$, 1h$_{11/2}$, and 1i$_{13/2}$ ...,  
respectively, appear due to the large spin-orbit splitting.  
Those high-j orbits have the parity different from other (normal) orbits 
in respective major shells.   Nevertheless, since the angular momentum 
of the high-j 
orbit differs by more than (or equal to) 2$\hbar$ from normal-parity orbits 
belonging to the same major shell, $E1$ transitions do not occur  
between the one-particle orbits belonging to a given major shell 
of the spherical 
nuclear potential.  Essentially the same mechanism continues to 
hinder strong low-frequency 
$E1$ transitions in quadrupole-deformed nuclei. 
Namely, one can hardly find an energetically close-lying pair of 
one-particle levels, of which the asymptotic quantum-numbers 
[$N \, n_z \, \Lambda \, \Omega$] allow $E1$ transitions between 
them.   

In contrast, in nuclei with weakly bound neutrons the origin of the hindrance 
(c) may be removed due to the change of the shell structure or 
the one-particle wave-functions, while the hindrance (b) may be 
drastically reduced due to the very weak coupling of weakly-bound neutrons 
with the well-bound core.   

In Ref. \cite{CL18} the lower limit of 0.030 e$^2$ fm$^2$ 
for the $B(E1; \, 1/2^+ \rightarrow 3/2^-)$ value in $^{27}$Ne was reported,  
which represented one of the strongest $E1$ strengths observed 
among the bound discrete states in this mass region.   It was also reported 
that the $E1$ transition strength is at least 30 times larger than 
the one measured for the decay of the 3/2$^-$ state 
to the 3/2$^+$ ground state.   
In the present paper I show an estimate of the $E1$ strength between 
the two excited states with weakly-bound neutrons based on 
the deformed Woods-Saxon potential, 
taking explicitly into account both the deformation and the feature of 
the wave functions of weakly-bound neutrons in the deformed finite potential. 
On the other hand, in Ref. \cite{CL18} the result of large-scale shell-model 
calculations was included, which could reproduce the observed order of 
magnitude of the $E1$ strength depending on effective interactions 
by using the harmonic oscillator basis in the calculation of radial integrals.  

Based on the measured $B(E2;0^{+}_{gr} \rightarrow 2^+_{1})$ values 
obtained in the analysis of heavy-ion inelastic scattering experiments  
\cite{JG07,HI05} it is often stated that even-even nuclei next to 
$^{27}$Ne, $^{26}$Ne and $^{28}$Ne, are deformed.
However, I note that the measured $E(2_1^+)$ value systematically varies 
as 2.018, 1.304 and 0.792 
MeV in $^{26}$Ne, $^{28}$Ne and $^{30}$Ne, respectively, 
and the measured energy ratio $E(4_1^+)/E(2_1^+)$ is 2.308 and 2.822 
in $^{28}$Ne and $^{30}$Ne, respectively.  
These data suggest that the transition from spherical to deformed shape takes
place in the isotope chain of $^{26}$Ne to $^{30}$Ne.   
Therefore, I expect that the nucleus $^{27}$Ne lies on the border of 
deformed and spherical nuclei. 
Then, it is likely that both deformed and spherical states appear 
in the region of low excitation energy of $^{27}$Ne.  
I suppose the ground state of $^{27}$Ne is basically spherical 
since the spin-parity of $3/2^{+}$ is naturally expected 
for the neutron number N=17 in spherical shape. 
On the other hand, the almost degenerate 1/2$^+$ and 3/2$^-$ levels imply that
these states are deformed.   If I consider one-particle levels in the deformed
potential this degeneracy is naturally understood, and as explained later, 
I can
estimate the deformations of these states as $\beta \approx 0.4$ based on the
deformed Woods-Saxon potential.  This estimate reasonably coincides with the
values estimated in Refs. \cite{JG07, HI05}.
Therefore, in the present work I concentrate on the estimate of 
the $B(E1; \, 1/2^+ \rightarrow 3/2^-)$ value assuming a prolate deformation 
for both excited states.   
I note that since the neutron separation energy $S_n$ of the ground state 
of $^{27}$Ne is 1.515 MeV, the $S_n$ values 
of the excited states 1/2$^+$ and 3/2$^-$ are  
only 0.63 and 0.75 MeV, respectively.  Then, in the estimate of 
$E1$ matrix-elements the radial part of relevant one-particle 
wave-functions has to be carefully prepared when the states have 
such small $S_n$ values.   

In Sec. II the main points of the model used are briefly summarized,   
and the results of numerical calculations in the present case of $^{27}$Ne 
are given, 
while in Sec. III the $E1$ transition between
well-bound neutron [200 1/2] and [3301/2] orbits is calculated 
by using the present model and is compared 
with the result in Sec. II.       
Conclusion and discussions are given in Sec. IV.

\section{MODEL AND NUMERICAL CALCULATIONS}
One-particle energy spectra in the quadrupole-deformed potential are known to 
have a close relation to the observed low-lying energy spectra 
in odd-A deformed nuclei \cite{BM75}.  In other words, 
the one-particle picture
in deformed nuclei seems to work much better than that in spherical nuclei.   
This is certainly because in the
phenomenological quadrupole-deformed potential the major part of the residual
interaction in the spherical potential, quadrupole-quadrupole interaction, is
already included in the one-body potential.  
Spectroscopic properties of low-energy levels of light deformed odd-N 
(or odd-Z) nuclei have been successfully analyzed based on 
deformed one-particle 
potential. For example, see the successful analysis of 
the spectroscopic properties of $^{25}$Mg and $^{25}$Al 
in Ref. \cite{BM75} taking the stable nucleus $^{24}$Mg as the deformed core. 
In Ref. \cite{IHS07} the data on the deformed nucleus $^{11}$Be were analyzed 
in terms of the model similar to the one in the present article.   However, 
one may wonder whether it is appropriate to use one-body potential 
in the description of such a light nucleus as $^{11}$Be.   
In the present work I apply the deformed one-body potential 
to the nucleus $^{27}$Ne with weakly-bound neutrons, 
knowing the successful work
on the well-bound neutrons in $^{25}$Mg \cite{BM75}.      

In the case of $^{27}$Ne the ground state (3/2$^{+}$) may be 
basically a spherical state, though  
in the 17th neutron wave-function the components 
other than $2d_{3/2}$ may not be 
negligible.  (The spectroscopic factor with rather large experimental errors
obtained from a d($^{26}$Ne, $^{27}$Ne)p reaction can be found, for example, 
in Ref. \cite{SMB12}.)
In contrast, I take the model that both the first (3/2$^-$) and 
second (1/2$^+$) excited states as the deformed $^{26}$Ne core 
plus one neutron 
in the quadrupole-deformed potential.  This is partly because 
the coexistence of spherical and deformed shape is expected in the region of 
low-excitation energy of $^{27}$Ne as described in Introduction and 
partly because in one-particle spectra of 
the prolately deformed potential in Fig. 1  
it is so easily expected that 
in nuclei with N=17 the deformed $3/2^{-}$ and $1/2^{+}$ states, 
which are almost degenerate, are energetically close-lying to 
a spherical $3/2^{+}$ state.  

As a spherical one-body potential I take the Woods-Saxon potential 
described in p.238-240 of Ref. \cite{BM69}.  Then, I examine the shell
structure of one-particle energy spectra 
in the axially-symmetric quadrupole-deformed potential 
which is constructed in a standard way.  
I write the single-particle wave-function in the body-fixed (intrinsic) 
coordinate system as 
\begin{equation}
\Psi_{\Omega}(\vec{r}) = \frac{1}{r} \, \sum_{\ell j} R_{\ell j  \Omega}(r) \,  
{\bf Y}_{\ell j \Omega}(\hat{r}),
\label{eq:twf}
\end{equation}
which satisfies 
\begin{equation}
H \, \Psi_{\Omega} = \varepsilon_{\Omega} \, \Psi_{\Omega}   
\label{eq:sheq}
\end{equation} 
where $\Omega$ expresses the component of one-particle angular-momentum 
along the symmetry axis and is a good quantum number.   
The Hamiltonian $H$ in Eq. (\ref{eq:sheq}) consists of the kinetic energy part 
and the potential energy part. The axially symmetric quadrupole-deformed 
potential 
for neutrons consists of the following three parts.  
\begin{equation}
V(r) = V_{WS} \, f(r), 
\label{eq:VWS}
\end{equation}
\begin{equation}
V_{coupl}(\vec{r}) = -\beta k(r) Y_{20}(\hat{r}), 
\label{eq:vcoupl}
\end{equation}
\begin{equation}
V_{so}(r) = - V_{WS} \, v \, \left(\frac{\Lambda}{2} \right)^{2} \, 
\frac{1}{r} \, \frac{df(r)}{dr} \, 
(\vec{\sigma} \cdot \vec{\ell}),  
\label{eq:eles}
\end{equation}
where  $\Lambda$ is the reduced Compton wave-length of nucleon 
$\hbar$/$m_{r} c$, 
\begin{equation}
f(r) = \frac{1}{1+exp \left(\frac{r-R}{a} \right)}
\label{eq:fr}
\end{equation}
and 
\begin{equation}
k(r) = R V_{WS} \frac{df(r)}{dr} \, . 
\label{eq:kr}
\end{equation} 
The values of the parameters in the potential 
except the depth $V_{WS}$ are taken from Ref. \cite{BM69}.   
For example, the diffuseness $a$ = 0.67 fm and 
the radius $R$= $r_{0} A^{1/3}$ with 
$r_0$ = 1.27 fm.  Taking the value $v$ = 32 in (\ref{eq:eles}) gives 
the same spin-orbit potential in Ref. \cite{BM69}.                 

The coupled differential equations for the radial wave-functions 
are written as 
\begin{equation}
\left(\frac{d^2}{dr^2} - \frac{\ell (\ell +1)}{r^2} + \frac{2m}{\hbar^2}( 
\varepsilon_{\Omega} - V(r) - V_{so}(r) ) \right) R_{\ell j \Omega}(r) = 
\frac{2m}{\hbar^2} 
\sum_{\ell^{'} j^{'}} \langle {\bf Y}_{\ell j \Omega} \mid V_{coupl} 
\mid {\bf Y}_{\ell^{'} j^{'}
\Omega} \rangle R_{\ell^{'} j^{'} \Omega}(r)
\label{eq:cpl}
\end{equation}
where 
\begin{eqnarray}
\langle {\bf Y}_{\ell j \Omega} \mid V_{coupl} \mid {\bf Y}_{\ell^{'} j^{'} 
\Omega} 
\rangle 
 & = &
- \beta \, k(r) \, \langle {\bf Y}_{\ell j \Omega} \mid Y_{20}(\hat{r}) \mid 
{\bf Y}_{\ell^{'} j^{'} \Omega} \rangle \nonumber \\
& = & 
- \beta \, k(r) \, (-1)^{\Omega-1/2} \, 
\sqrt{\frac{(2j+1)(2j^{'}+1)}{20 \pi}} \nonumber \\
&& \mbox{ } 
C(j, j^{'}, 2;\Omega, -\Omega, 0) \, C(j, j^{'}, 2; \frac{1}{2}, -\frac{1}{2}, 
0).   
\end{eqnarray} 

The details of the model 
can be found, for example, 
in Ref.  \cite{IH19} and the quoted papers therein.  
The important point in the present work 
is that the coupled differential equations are integrated 
in coordinate space with the correct asymptotic behavior of radial 
wave functions in respective $(\ell,j)$ channels for $r \rightarrow \infty$.   
In particular, the r-dependence of a given $(\ell, j)$ component 
thus obtained is generally different from that of eigenfunctions 
in the spherically symmetric potential due to the presence 
of the coupling term with the deformed field on the right-hand-side of 
Eq. (\ref{eq:cpl}), which is proportional 
to $\beta $ .   
Such a unique r-dependence of the solution of 
Eq. (\ref{eq:cpl}) is clearly seen, for example, in   
the d$_{5/2}$ component of the  [200 1/2] orbit shown in Fig. 5.  
Some detailed  
discussion related to this point is given in the last paragraph of Sec. III.  
In this way, the values 
of the radial matrix-element of $E1$ operator, $\langle r \rangle$, 
between weakly-bound one-particle wave-functions in the deformed potential are  
carefully calculated including the possible halo effect. 

For axially symmetric quadrupole-deformed shape an infinite number of channels 
with ($\ell, j, \Omega$) are coupled for a given $\Omega$.  In practice, 
the number of coupled channels or the maximum value of ($\ell, j$) pairs 
included for a given $\Omega$ is determined so that 
the resulting eigenvalues and 
wave functions do not change in a meaningful way by taking a larger number of 
coupled channels.  In the present work 
s$_{1/2}$, d$_{3/2}$, d$_{5/2}$, g$_{7/2}$, 
and g$_{9/2}$ are included in the calculation of the [200 1/2] orbit, 
while in the calculation of the [330 1/2] orbit 
p$_{1/2}$, p$_{3/2}$, f$_{5/2}$, f$_{7/2}$, h$_{9/2}$, and h$_{11/2}$ 
are included.   It is noted that the ($\ell, j, \Omega$) component obtained 
by integrating the coupled differential equations (\ref{eq:cpl}) in the present model 
contains properly the contributions from ($n, \ell, j, \Omega$) orbits 
with various radial nodes $n$.    

In Fig. 1 I show the calculated one-neutron energy eigenvalues in the deformed 
Woods-Saxon potential produced by the $^{26}$Ne core   
as a function of quadrupole deformation parameter $\beta$.   All parameters of
the potential are taken from those given by Ref. \cite{BM69} except the depth, 
$V_{WS}$ = $-$41.0 MeV, which was used so that 
the crossing energy of the levels [220 1/2] and [330 1/2] is 
around $\varepsilon_{\Omega} = -$650 keV.  
The value of 650 keV is chosen to be approximately equal to 
the observed binding
energies of the neutron in the 1/2$^{+}$ and 3/2$^{-}$ states.  
The two levels cross 
at $\beta = 0.44$ in the case of Fig. 1.  
If $V_{WS}$ = $-$43 MeV is used the two levels cross at $\beta$ = 0.42.   
In both cases the energies of two levels at the crossing point are always 
close to the energy of 1d$_{3/2}$ at $\beta$ = 0.  This is consistent with the
assumption of the spherical ground state and deformed excited states, which I
made in this study.  
The band-head states generated by the [200 1/2] and [330 1/2] configurations
should be  
1/2$^{+}$ and 3/2$^{-}$, respectively.   The latter can be understood  
from the fact that
the main components of 
the [330 1/2] state are $1f_{7/2}$ and $2p_{3/2}$, therefore, 
the decoupling parameter of the 
[330 1/2] band is between $-$4 (for f$_{7/2}$) and $-$2 (for p$_{3/2}$).   \
Consequently, the angular momentum of the band-head state of 
the [330 1/2] configuration  
is not equal to $I = 1/2$, but 3/2.   

The formula for $B(E1)$ in the present case is written as 
(see eq. (4-91) in Ref. \cite{BM75} for the reduced matrix-elements)  
\begin{eqnarray}
B(E1&;&I^{\pi}=K^{\pi}=1/2^{+} \rightarrow I^{\pi}=3/2^{-}, K^{\pi}=1/2^{-}) 
 \nonumber \\
& = &  (e_{eff}^{n}(E1))^2 \, 
\{ C(\frac{1}{2},1,\frac{3}{2};\frac{1}{2},0,\frac{1}{2}) \,
\langle \, [330 \, 1/2] \,| rY_{10} | \, [200 \, 1/2] \, \rangle \nonumber \\
 && \mbox{ } + (-1)^{1/2+1/2} \, C(\frac{1}{2},1,
\frac{3}{2}; \frac{-1}{2},1, \frac{1}{2} ) \, 
\langle \, [330 \, 1/2] \, | rY_{11} | \, [\widetilde{200 \, 1/2}] \, 
\rangle \}^2 
\label{eq:BE1} 
\end{eqnarray}
where $[\widetilde{200 \, 1/2}]$ expresses the time-reversed intrinsic
configuration of [200 1/2].     
By using one-particle wave functions of the [330 1/2] and [200 1/2] states 
calculated for $V_{WS} = -41.0$ MeV and $\beta$ = 0.44 I obtain 
\begin{eqnarray}
\langle \, [330 \, 1/2] \, | rY_{10} | \, [200 \, 1/2] \, \rangle & = & 0.458 
\quad \mbox{fm} \nonumber \\
\langle \, [330 \, 1/2] \, | rY_{11} | \, [\widetilde{200 \, 1/2}] \, 
\rangle & = & -0.013 \quad \mbox{fm} 
\label{eq:rY1}
\end{eqnarray} 
where the smallness of the $rY_{11}$ matrix elements is the result of 
cancellation between major contributions.  
Using the values of (\ref{eq:rY1}) in (\ref{eq:BE1}), I obtain 
\begin{equation}
B(E1;I^{\pi}=K^{\pi}=1/2^{+} \rightarrow I^{\pi}=3/2^{-}, K^{\pi}=1/2^{-}) 
\, = \, [e^{n}_{eff}(E1)]^2 \, (0.145) \quad \mbox{e$^2$ fm$^2$}
\label{eq:BE1n}
\end{equation} 

In Figs. 2 and 3 the components of the radial wave functions, 
$R_{\ell j \Omega}(r)$ of the [200 1/2] 
and [330 1/2] states, respectively, are shown as a function of 
radial coordinate.  
It is seen that due to the small binding energies the major part 
of the wave functions especially with smaller orbital angular-momentum 
lies outside the Woods-Saxon potential.   
Namely, the large halo phenomena, in particular in the orbits $s_{1/2}$ 
in Fig. 2 and $p_{3/2}$ and $p_{1/2}$ in Fig. 3, are recognized.     

In all $\Omega^{\pi}$ = 1/2$^{+}$ orbits the probability 
of the $s_{1/2}$ component 
becomes unity in the limit of 
$\varepsilon_{\Omega} \rightarrow 0$  \cite{TM97, IH04}.  
However, at the binding energy
of 622 keV in Fig. 2 I do not find the tendency that 
the probability 
of the s$_{1/2}$ component  
in the [200 1/2] orbit is particularly large. 
The probability of the  
s$_{1/2}$ component in Fig. 2 is 0.518 compared with 0.492 
in the well-bound case of Fig. 5.   
The binding
energy at which the probability of s$_{1/2}$ component starts to steeply
increase depends very much on individual orbits.  See Fig. 3 of Ref. \cite{IH04}.  
In the case of the [200 1/2] level, which is the energetically highest 
$\Omega^{\pi} = 1/2^{+}$ level for prolate shape that comes from the sd-shell, 
the steep increase of the s$_{1/2}$ probability starts 
first for $|\varepsilon_{\Omega}| \simleq 500$ keV.     

In Table I the respective contributions 
$\langle \ell_1 \,  j_1 \,  \Omega_1^{\pi}=1/2^- \, | rY_{10} | \, 
\ell_2 \,  j_2 \,  
\Omega_2^{\pi}=1/2^+ \rangle$  
to the matrix element in Eq. (\ref{eq:rY1}) 
\begin{equation}
\langle \, [330 \, 1/2]\, | \, rY_{10} \, | \, [200 \, 1/2] \, \rangle \, = 
\, \sum_{\ell_1, j_1, \ell_2, j_2} \langle \, \ell_1 \, j_1 \, 
\Omega_1^{\pi}=1/2^- 
\, | rY_{10} | \, \ell_2 \, j_2 \, \Omega_2^{\pi}=1/2^+ \, \rangle
\label{eq:comp}
\end{equation} 
are shown.   
It is seen that the large positive contribution comes
from the two transition matrix-elements between particular halo components, 
$s_{1/2} \rightarrow p_{1/2}$ and 
$s_{1/2} \rightarrow p_{3/2}$, while almost all other contributions 
that are much
smaller have different (negative) sign and, thus, 
cancel partially the halo contribution.  
Halo components hardly polarize the well-bound core, while the latter 
(negative) contributions may be considerably affected by the polarization of 
the core.   Consequently, if the dependence of the $e_{eff}^{n}(E1)$ value on
one-particle wave-functions is taken into account, the value 
$|e_{eff}^{n}(E1)|$ = (Z/A)$e$ may be the first approximation for halo
contributions, while other (negative) contributions may be associated with 
the $|e_{eff}^{n}(E1)|$ value 
which is considerably smaller than (Z/A)$e$.   
Then, the magnitude of the cancellation of the $E1$ matrix-element coming from 
the halo components by non-halo components may become smaller.   
However, since 
the detailed dependence of $e_{eff}^{n}$ on one-particle wave-functions 
lies outside the scope of the present paper, here I use  
$|e_{eff}^{n}(E1)|$ = (Z/A)$e$ for all contributions, as an estimate of the 
$B(E1)$ value.  Then, I obtain 
$B(E1;I^{\pi}=K^{\pi}=1/2^{+} \rightarrow I^{\pi}=3/2^{-}, K^{\pi}=1/2^{-}) $ 
= 0.020 e$^2$ fm$^2$, which is the order of magnitude of the measured $B(E1)$ 
value.   
Since there are several adjustable parameters in the present model 
such as the radius and diffuseness parameters 
of the deformed Woods-Saxon potential and 
the dependence of $e_{eff}^{n}(E1)$ value on one-particle orbits, 
here I am satisfied with the agreement of the order of magnitude.  
However, one has to keep in mind that 
the value of $e_{eff}(E1)$ is 
the important 
parameter which really decides calculated $B(E1)$ values, though it is often 
carelessly chosen.    

From Table I it is seen that, for example, the $f_{7/2}$ component 
has 40 percent probability in the one-particle wave-function of [330 1/2], 
nevertheless, it has only a negligible contribution     
to the present $E1$ transition matrix-element. 

\section{E1 TRANSITION BETWEEN WELL-BOUND NEUTRON ORBITS, [200 1/2] 
AND [330 1/2]}
In order to understand the characteristic feature 
of the enhanced $E1$ transition  
between the weakly-bound neutron orbits, [200 1/2] and [330 1/2], 
which is obtained 
in the previous section,  
in this section  
the $E1$ matrix-element is calculated in the case that both orbits are 
well-bound, 
say, 
bound by about 8 MeV, using the same parameters of the Woods-Saxon potential 
except the depth.      

For the depth $V_{WS}$ = $-$ 55.81 MeV the 1d$_{3/2}$ orbit 
in the spherical potential 
($\beta$=0) is bound by $-$7.99 MeV, which is roughly the energy  
of the least-bound neutron 
in $\beta$-stable nuclei.    
In Fig. 4 
calculated neutron one-particle energies in the potential are plotted 
as a function of quadrupole deformation parameter $\beta$.   
It is seen that the [200 1/2] and [330 1/2] orbits cross with each other 
again 
at the energy which is almost equal to that of the 1d$_{3/2}$ level 
at $\beta$ = 0.  
The deformation parameter at the level crossing is $\beta$ = 0.30 
for this potential depth.     However, I am interested 
in the $E1$ matrix element between the [200 1/2] and [330 1/2] 
orbits at $\beta$ = 0.44 that is the deformation, 
for which the enhanced $E1$ matrix element was 
obtained in Sec. II.   For $\beta$ = 0.44 I obtain 
\begin{eqnarray}
\langle \, [330 \, 1/2] \, | rY_{10} | \, [200 \, 1/2] \, \rangle & = & 0.035 
\quad \mbox{fm} \nonumber \\
\langle \, [330 \, 1/2] \, | rY_{11} | \, [\widetilde{200 \, 1/2}] \, 
\rangle & = & 0.000 \quad \mbox{fm} 
\label{eq:rY1-app}
\end{eqnarray} 
Then, using the values of (\ref{eq:rY1-app}) in (\ref{eq:BE1}), I obtain 
\begin{equation}
B(E1;I^{\pi}=K^{\pi}=1/2^{+} \rightarrow I^{\pi}=3/2^{-}, K^{\pi}=1/2^{-}) 
\, = \, [e^{n}_{eff}(E1)]^2 \, (0.00082) \quad \mbox{e$^2$ fm$^2$}
\label{eq:BE1n-app}
\end{equation} 
When I use 
$|e_{eff}^{n}(E1)|$ = (Z/A)$e$ for all contributions, I obtain 
$B(E1;I^{\pi}=K^{\pi}=1/2^{+} \rightarrow I^{\pi}=3/2^{-}, K^{\pi}=1/2^{-}) 
$ 
= 0.00011 e$^2$ fm$^2$, which is more than two orders of magnitude 
smaller than the $B(E1)$ 
value, 0.020 e$^{2}$ fm$^{2}$, obtained for weakly-bound neutrons in Sec. II.   

In Table II the respective contributions 
$\langle \ell_1 \,  j_1 \,  \Omega_1^{\pi}=1/2^- \, | rY_{10} | \, 
\ell_2 \,  j_2 \,  
\Omega_2^{\pi}=1/2^+ \rangle$  
to the matrix element in Eq. (\ref{eq:rY1-app}) 
are shown.   
It is seen that in Table II ; (i) the positive contributions 
from $s_{1/2} \rightarrow p_{3/2}$ 
and $s_{1/2} \rightarrow p_{1/2}$ to the matrix element of $rY_{10}$ are 
a factor of 3 to 4 smaller than the corresponding contributions in Table I 
and (ii) 
the cancellation of the positive contributions 
by other negative contributions is efficient.   
The result (i) comes partly from the halo effect of the radial wave functions 
of the weakly-bound s and p neutrons in the case of Table I and partly 
from the larger 
probability (0.512)
of the p$_{3/2}$ component in the weakly-bound [330 1/2] orbit   
in the case of Table I compared 
with the probability (0.299) of the p$_{3/2}$ component 
in the well-bound [330 1/2] orbit in Table II.  
The much larger probability of the p$_{3/2}$ component 
in the weakly-bound [330 1/2] 
orbit is mainly the result of  
the unique shell-structure that when both 2p$_{3/2}$ and 1f$_{7/2}$ neutron orbits 
are weakly bound or one-particle resonant the 2p$_{3/2}$ level 
is almost degenerate with or 
slightly lower than the 1f$_{7/2}$ level \cite{IH07,IH10,IH12}.    
In contrast, in the case of Fig. 4 
the 1f$_{7/2}$ level lies lower than the 2p$_{3/2}$ level by 2.4 MeV, 
therefore, for moderately-deformed prolate shape f$_{7/2}$ is 
the main component of the lowest $\Omega^{\pi}$ = 1/2$^{-}$ level 
in the fp-shell 
denoted by [330 1/2].   
The strong cancellation (ii) is expected,  
since the $rY_{10}$ matrix element vanishes 
if the wave functions of the [200 1/2] and [330 1/2] orbits are exactly 
those specified 
by the respective asymptotic quantum numbers and since the wave functions 
of well-bound neutrons are more similar to those specified 
by asymptotic quantum numbers 
than those of  
weakly-bound neutrons. 

From the analysis of numerical results which is described above 
and in the previous section 
it is understood that the enhanced $E1$ transition 
between weakly-bound excited states of $^{27}$Ne 
obtained  
in the present work comes from the halo behavior 
of s and p components of the wave functions of the weakly-bound neutron  
[200 1/2] and [330 1/2] orbits, in addition to the shell-structure 
unique in weakly-bound neutrons.

The r-dependence of the d$_{5/2}$ component in Fig. 5 is 
qualitatively different 
from that of eigenfunctions of the spherically symmetric potential 
such as 1d$_{5/2}$ 
and 2d$_{5/2}$.    
In the following I consider the case of moderate deformation. 
If the coupling term on the right-hand-side 
of Eq. (\ref{eq:cpl}) is 
weak enough to be treated by perturbation in the ($\ell j \Omega$) channel, 
the r-dependence 
of $R_{\ell j \Omega}(r)$ is, roughly speaking, similar to 
that of the energetically closest-lying 
(namely, $\varepsilon_{\Omega} \approx \varepsilon_{n \ell j}$) 
eigenfunction with ($n, \ell, j$) of 
the spherically symmetric potential.    
This is usually the case that 
the $(\ell j \Omega)$ component has a considerable probability 
in the wave function of the one-particle orbit with $\varepsilon_{\Omega}$.
In contrast, if the energy $\varepsilon_{\Omega}$ is far away 
from any $\varepsilon_{n \ell j}$  for a given ($\ell, j$) 
the probability of the ($\ell j \Omega$) component in the one-particle orbit 
with 
$\varepsilon_{\Omega}$ is usually small.  Then, when the radial wave function 
$R_{\ell j \Omega}(r)$ is expanded 
in terms of eigenfunctions with ($n, \ell, j$)
in the spherically symmetric potential, namely in terms of the solutions 
of Eq. (\ref{eq:cpl}) 
in the absence of the coupling term on the right-hand-side, 
the major components 
of $R_{\ell j \Omega}(r)$ will be generally 
those with ($n, \ell, j, \Omega$), 
of which the energies $\varepsilon(n \ell j)$ are closer 
to $\varepsilon_{\Omega}$.  
In the case of Fig. 5 the probability of $d_{5/2}$ component 
in the [200 1/2] orbit is small (0.039),   
and the energy of the [200 1/2] orbit, $\varepsilon$([200 1/2]), 
is between $\varepsilon$(1d$_{5/2}$) and $\varepsilon$(2d$_{5/2}$),
but closer to $\varepsilon$(1d$_{5/2}$).   Consequently, 
the r-dependence of the d$_{5/2}$ component in the [200 1/2] orbit shows 
something 
between that of 1d$_{5/2}$ and 2d$_{5/2}$ orbits and, consequently, 
has a node at r = 5.2 fm 
that is outside the nuclear radius.   In the case that 
$\varepsilon$([200 1/2]) is closer to $\varepsilon$(2d$_{5/2})$, 
the node moves 
towards the inside of nuclear radius, and the radial shape 
of the d$_{5/2}$ component 
in the [200 1/2] orbit will become qualitatively similar to 
that of 2d$_{5/2}$ orbit.          Though the unique r-dependence of 
some solutions of Eq. (\ref{eq:cpl}) is interesting, 
it is not playing a crucial role 
in the enhanced E1 transition presently calculated.   
However, the  
one-particle wave functions, 
which have the r-dependence very different from that of any eigenfunctions of 
spherically symmetric potential, may produce, for example, 
cross sections of one-particle transfer reactions, 
which are unexpected from the traditional analysis of the reactions 
using one-particle wave functions 
with ($n, \ell, j$) quantum numbers \cite{IH74}.

\section{CONCLUSION AND DISCUSSIONS}
Interpreting that the first and second excited states of $^{27}$Ne, 
3/2$^-$ at Ex=0.765 MeV and 1/2$^+$ at Ex=0.885 MeV, 
are prolately deformed around 
$\beta \approx 0.4$, the order of magnitude of the observed enhanced 
$B(E1; 1/2^+ \rightarrow 3/2^-)$ value is obtained by the present model.  
The large $B(E1)$ value comes from the halo behavior of 
the $s_{1/2}$ component in the [200 1/2] orbit as well as the $p_{1/2}$ 
and $p_{3/2}$ components in the [330 1/2] orbit, 
in addition to the large probability (0.512) 
of the p$_{3/2}$ component in the [330 1/2] orbit which is even larger than 
the probability (0.399) of the f$_{7/2}$ component. 
The large probability of the p$_{3/2}$ component is 
originated from the near degeneracy 
of the 2p$_{3/2}$ and 1f$_{7/2}$ levels at $\beta$ = 0, 
which is the shell
structure unique in the very-weakly
bound or resonant one-neutron levels.  
In this respect, it should be noted that the $E1$ matrix-element 
between the [200 1/2] and [330 1/2] 
orbits is in fact vanishing if the wave functions of the two orbits    
are those exactly specified by the respective asymptotic quantum-numbers, 
[$N \, n_z \, \Lambda \, \Omega$], since the $rY_{1 \mu}$ operator 
cannot change $n_z$ by 3.         

I also note that the occurrence of almost degenerate low-lying states 
with different parity, 1/2$^{+}$ and 3/2$^-$, is simply the result of 
deformation. Furthermore, as seen from Fig. 1, those two excited states 
are expected to be found very close to the basically spherical 
$3/2^+$ ground-state, in agreement with the observed spectra.   

The spin-parity  3/2$^+$ of the ground state of $^{27}$Ne is 
easily expected for an almost spherical nucleus with the neutron number N=17.  
On the other hand, 
since the shape coexistence is likely in the low-energy region  
of $^{27}$Ne, the lowest-lying levels of 3/2$^{-}$ from the [330 1/2] orbit 
and 1/2$^{+}$ from the [200 1/2] orbit are naturally expected in the region of
low-excitation energy for prolate deformation, which is certainly favored 
by the proton number Z=10.   
It is eagerly hoped to obtain any experimental information on the shape of the
1/2$^+$ state at Ex=0.885 MeV and the 3/2$^{-}$ state at Ex=0.765 MeV 
of $^{27}$Ne.   

The interpretation that the ground state of $^{27}$Ne is basically spherical 
while the first and second excited states, 3/2$^{-}$ and 1/2$^{+}$
at 0.765 and 0.885 MeV, respectively, are deformed goes well 
with the experimental observation \cite{CL18} that 
the present $1/2^+ \rightarrow 3/2^{-}$ $E1$ transition rate 
is much larger (at least 30 times) than that measured for the $3/2^-$ 
decay to the 3/2$^+$ ground state.

\newpage

\begin{table}
\caption{\label{table1} 
The contributions $\langle \, \ell_1 \, j_1 \, \Omega_1^{\pi}=1/2^{-} \, 
|rY_{10}| \, \ell_2 \, j_2 \, \Omega_2^{\pi}=1/2^{+} \, \rangle$  
in the unit of fm to 
the matrix element $\langle \, [330 \, 1/2] \, | rY_{10} | \, 
[200 \, 1/2] \, \rangle$, using the wave functions shown in Figs. 2 and 3.  
See eq. (\ref{eq:comp}).
}
\begin{center}
\begin{tabular}{cc|cccccc}  \hline
  &  & \multicolumn{6}{c}{[330 1/2]}   \\ 
   &   & $p_{1/2}$  &  $p_{3/2}$  &  $f_{5/2}$  &  $f_{7/2}$  &  $h_{9/2}$ & 
   $h_{11/2}$  \\ \hline  
[200 1/2] \quad &  $s_{1/2}$  & 0.165  & 0.608 &   &   &   &   \\
  &  $d_{3/2}$  & $-$0.105 & $-$0.038  & $-$0.066 &  &  &   \\ 
  &  $d_{5/2}$  &   & 0.003 & $-$0.000 & $-$0.061  &   &    \\
  &  $g_{7/2}$  &   &   & $-$0.013 & $-$0.002 & $-$0.002 &    \\ 
  &  $g_{9/2}$  &  &  &  & $-$0.025 & $-$0.000  & $-$0.004    \\ \hline
\end{tabular}
\end{center}
\end{table}


\begin{table}
\caption{\label{table2} 
The contributions $\langle \, \ell_1 \, j_1 \, \Omega_1^{\pi}=1/2^{-} \, 
|rY_{10}| \, \ell_2 \, j_2 \, \Omega_2^{\pi}=1/2^{+} \, \rangle$  
in the unit of fm 
to the matrix element $\langle \, [330 \, 1/2] \, | rY_{10} | \, 
[200 \, 1/2] \, \rangle$, calculated by using the wave functions shown 
in Figs. 5 and 6.   
See eq. (\ref{eq:comp}).
}
\begin{center}
\begin{tabular}{cc|cccccc}  \hline
  &  & \multicolumn{6}{c}{[330 1/2]}   \\ 
   &   & $p_{1/2}$  &  $p_{3/2}$  &  $f_{5/2}$  &  $f_{7/2}$  &  $h_{9/2}$ & 
   $h_{11/2}$  \\ \hline  
[200 1/2] \quad &  $s_{1/2}$  & 0.044  & 0.197 &   &   &   &   \\
  &  $d_{3/2}$  & $-$0.024 & $-$0.007  & $-$0.044 &  &  &   \\ 
  &  $d_{5/2}$  &   & 0.021 & $-$0.000 & $-$0.082  &   &    \\
  &  $g_{7/2}$  &   &   & $-$0.011 & $-$0.003 & $-$0.002 &    \\ 
  &  $g_{9/2}$  &  &  &  & $-$0.043 & $-$0.000  & $-$0.009    \\ \hline
\end{tabular}
\end{center}
\end{table}


\newpage

\noindent
{\bf\large Figure captions}\\
\begin{description}
\item[{\rm Figure 1 :}]
Calculated neutron one-particle energies in the potential produced 
by $^{26}_{10}$Ne$_{16}$ as a function of quadrupole deformation 
parameter $\beta$.
Some neutron numbers, which are obtained by filling all lower-lying levels, 
are indicated with open circles.   One-particle bound and resonant energies 
at $\beta = 0$ are $-$6.155, $-$3.810, $-$0.650, and $+$3.241 MeV 
for the $1d_{5/2}$, $2s_{1/2}$, $1d_{3/2}$, and $1f_{7/2}$ levels, 
respectively.   
The one-particle resonant level of $2p_{3/2}$, which is expected 
below that of $1f_{7/2}$, is not obtained when the one-particle resonance 
is defined 
in terms of phase shift.   The asymptotic quantum-numbers, 
$[N n_z \Lambda \, \Omega]$, are denoted for the one-particle levels, 
which are 
of particular importance in the present subject.   
Some one-particle resonant levels ($\varepsilon_{\Omega} > 0$) 
for $\beta \neq 0$, which are defined 
in terms of the eigenphase $\delta_{\Omega}$ 
\cite{IH05},  
are not plotted if they are not relevant to the present interest. 
\end{description}

\begin{description}
\item[{\rm Figure 2 :}]
Components of the radial wave function $R_{\ell j \Omega=1/2}(r)$ 
of the neutron [200 1/2] orbit at $\beta$ = 0.44 in Fig. 1 are shown 
as a function of radial coordinate.  Respective $(\ell j)$ quantum numbers 
are denoted without writing the radial node quantum-number $n$, 
because the wave functions are not eigenfunctions 
of any spherical potential.  The probabilities of respective $(\ell j)$ 
components in the wave function are 0.518, 0.443, 0.024, and 0.013 for  
$s_{1/2}$, $d_{3/2}$, $d_{5/2}$, and $g_{7/2}$.  Since the probability of 
$g_{9/2}$ is only 0.002, the $g_{9/2}$ wave function is not plotted.   
The thin vertical line at $r$ = 3.76 fm denotes the radius of 
the Woods-Saxon potential used.
\end{description}

\begin{description}
\item[{\rm Figure 3 :}]
Components of the radial wave function $R_{\ell j \Omega=1/2}(r)$ 
of the neutron [330 1/2] orbit at $\beta$ = 0.44 in Fig. 1 are shown 
as a function of radial coordinate.  Respective $(\ell j)$ quantum numbers 
are denoted without writing the radial node quantum-number $n$, 
because the wave functions are not eigenfunctions 
of any spherical potential.  The probabilities of respective $(\ell j)$ 
components in the wave function are 0.068, 0.512, 0.010, 0.399 for  
$p_{1/2}$, $p_{3/2}$, $f_{5/2}$, and $f_{7/2}$.  Since the probabilities of 
$h_{9/2}$ and $h_{11/2}$ are only 0.000 and 0.011, respectively, 
the $h_{9/2}$ 
and $h_{11/2}$ wave functions are not plotted.   
The thin vertical line at $r$= 3.76 fm denotes the radius 
of the Woods-Saxon potential used.
\end{description}

\begin{description}
\item[{\rm Figure 4 :}]
Calculated neutron one-particle energies in the potential, 
of which the parameters are the same as those in Fig.1 except the depth 
of the Woods-Saxon potential, $V_{WS}$, as a function of quadrupole deformation 
parameter $\beta$.   
The neutron number N=16, which are obtained by filling all lower-lying levels, 
are indicated with open circles.   Bound one-particle energies 
at $\beta = 0$ are $-$16.11, $-$11.75, $-$7.99, $-$4.26, and $-$1.81 MeV 
for the $1d_{5/2}$, $2s_{1/2}$, $1d_{3/2}$, $1f_{7/2}$, and $2p_{3/2}$ levels, 
respectively.   
The asymptotic quantum-numbers, 
$[N n_z \Lambda \, \Omega]$, are denoted for the one-particle levels, 
which are of particular importance in the present subject.   
Only one-particle energies that are important for the discussion 
of the present subject are plotted, which are, for example, 
the N=17th neutron level and the position of the 2p$_{3/2}$ level 
relative to the 1f$_{7/2}$.   Consequently, the $\Omega^{\pi}$ = 5/2$^{-}$ 
and 
7/2$^{-}$ levels connected to the 1f$_{7/2}$ level at $\beta$=0 and 
$\Omega^{\pi}$ = 3/2$^{-}$ level connected to the 2p$_{3/2}$ level 
at $\beta$=0 are not 
plotted.    
\end{description}

\begin{description}
\item[{\rm Figure 5 :}]
Components of the radial wave function $R_{\ell j \Omega=1/2}(r)$ 
of the neutron [200 1/2] orbit at $\beta$ = 0.44 in Fig. 4 are shown 
as a function of radial coordinate.  Respective $(\ell j)$ quantum numbers 
are denoted without writing the radial node quantum-number $n$, 
because the wave functions are not eigenfunctions 
of any spherical potential.  The probabilities of respective $(\ell j)$ 
components in the wave function are 0.492, 0.449, 0.039, and 0.016 for  
$s_{1/2}$, $d_{3/2}$, $d_{5/2}$, and $g_{7/2}$.  Since the probability of 
$g_{9/2}$ is only 0.003, the $g_{9/2}$ wave function is not plotted.  
See the text for the understanding of the clearly peculiar r-dependence of the   
wave function of the d$_{5/2}$ component.     
The thin vertical line at $r$ = 3.76 fm denotes the radius of 
the Woods-Saxon potential used.
\end{description}

\begin{description}
\item[{\rm Figure 6 :}]
Components of the radial wave function $R_{\ell j \Omega=1/2}(r)$ 
of the neutron [330 1/2] orbit at $\beta$ = 0.44 in Fig. 4 are shown 
as a function of radial coordinate.  Respective $(\ell j)$ quantum numbers 
are denoted without writing the radial node quantum-number $n$, 
because the wave functions are not eigenfunctions 
of any spherical potential.  The probabilities of respective $(\ell j)$ 
components in the wave function are 0.021, 0.299, 0.007, 0.642 for  
$p_{1/2}$, $p_{3/2}$, $f_{5/2}$, and $f_{7/2}$.  Since the probabilities of 
$h_{9/2}$ and $h_{11/2}$ are only 0.000 and 0.029, respectively, 
the $h_{9/2}$   
and $h_{11/2}$ wave functions are not plotted.      
The thin vertical line at $r$= 3.76 fm denotes the radius 
of the Woods-Saxon potential used.
\end{description}

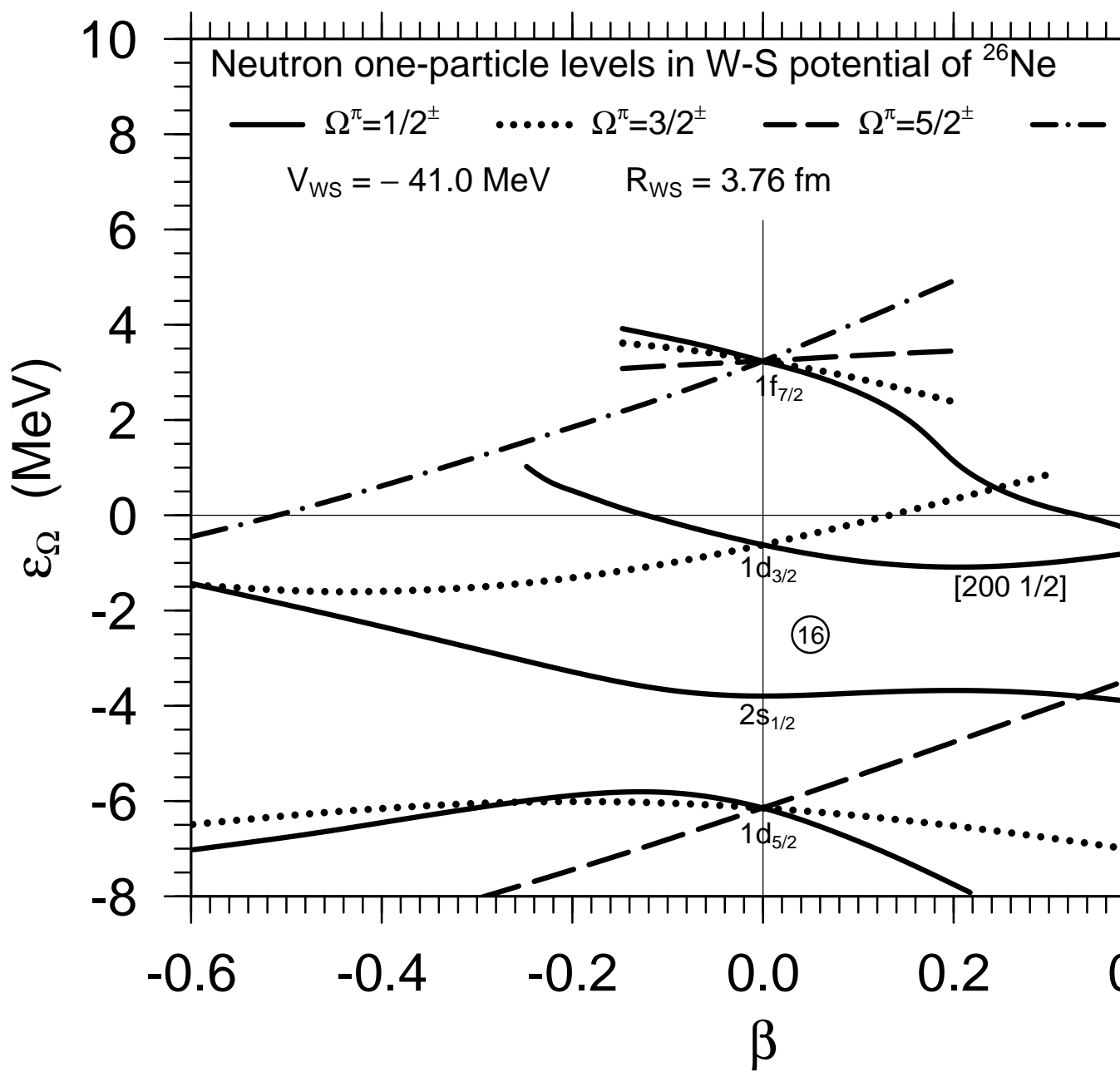
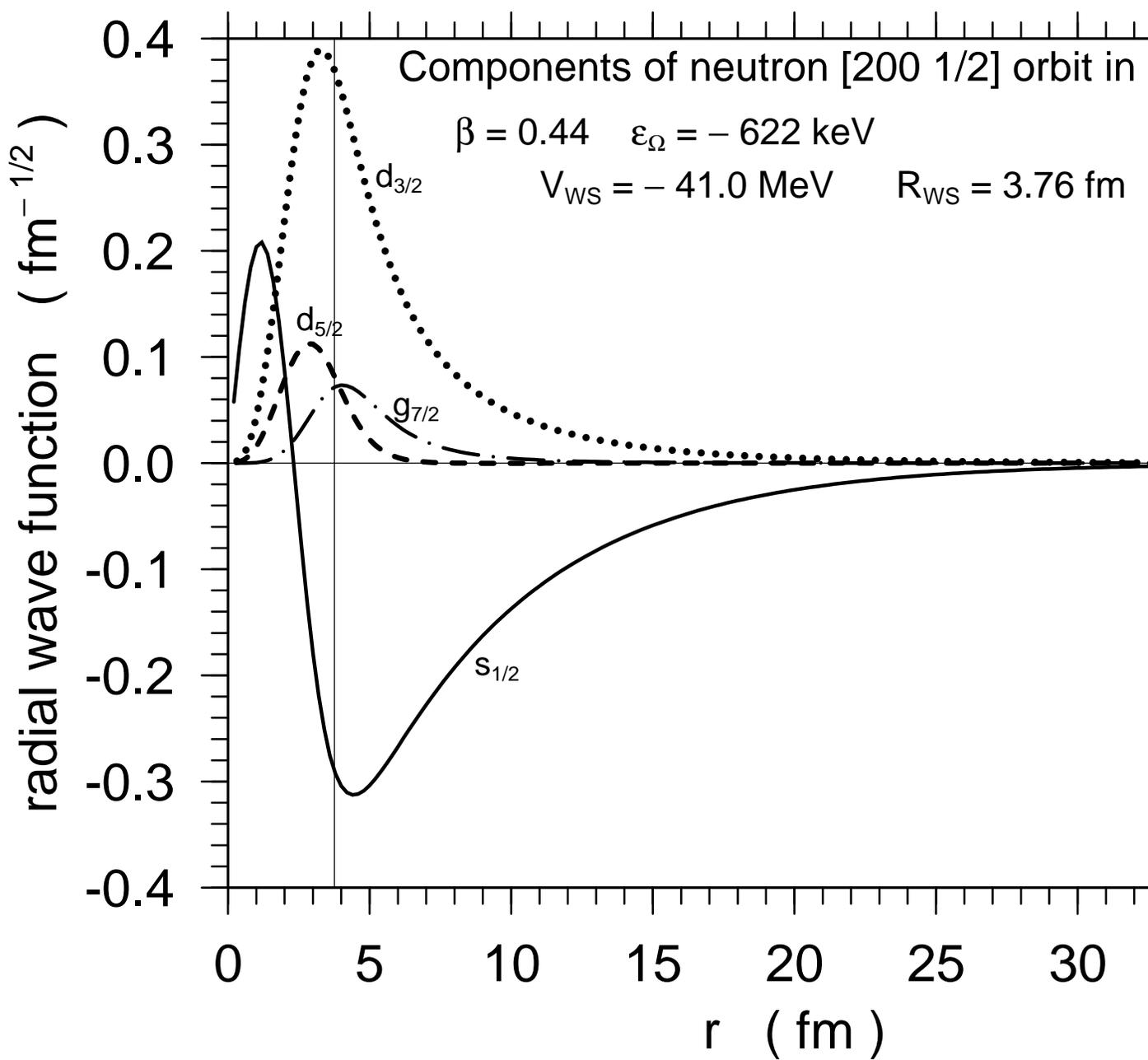
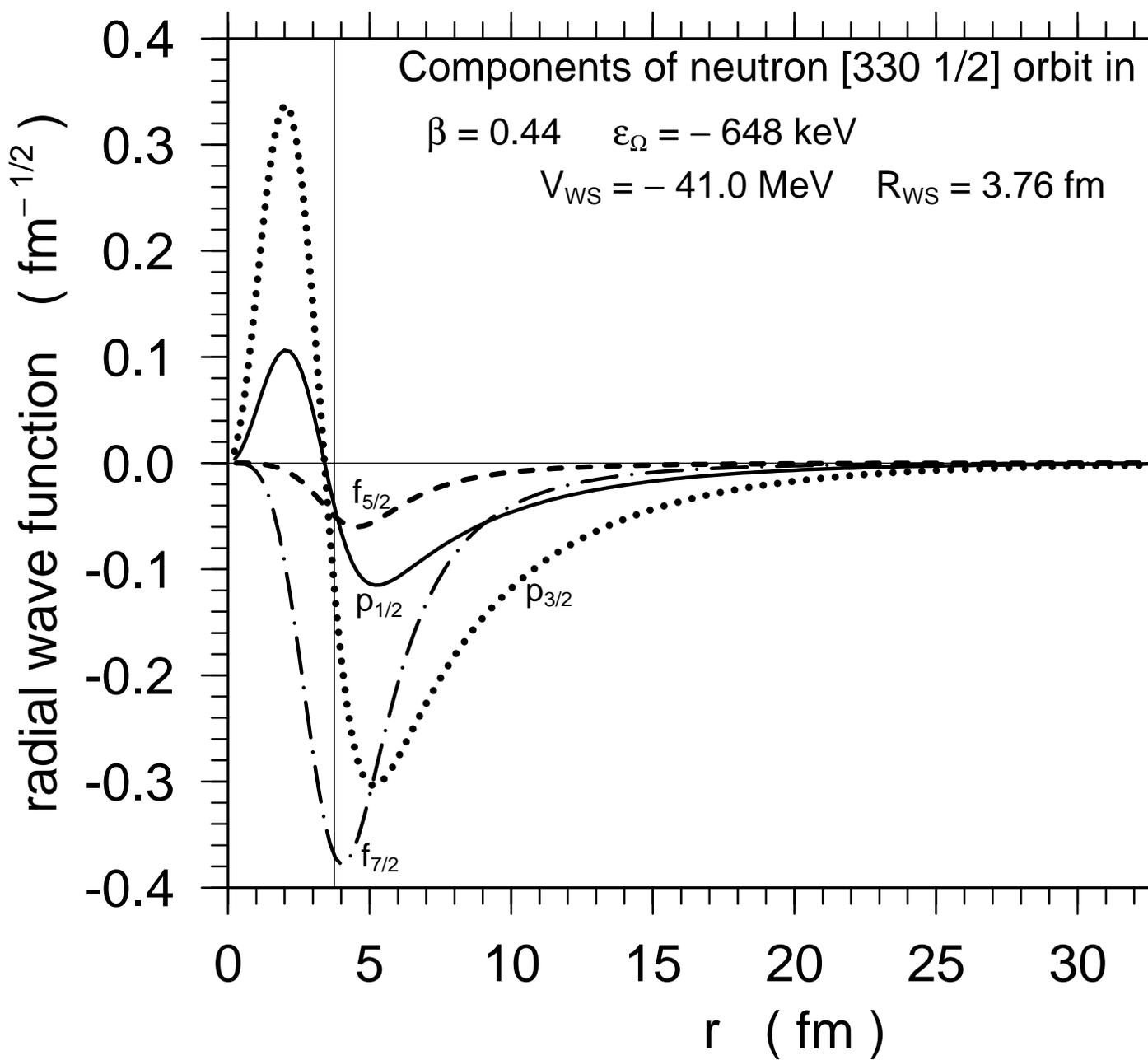
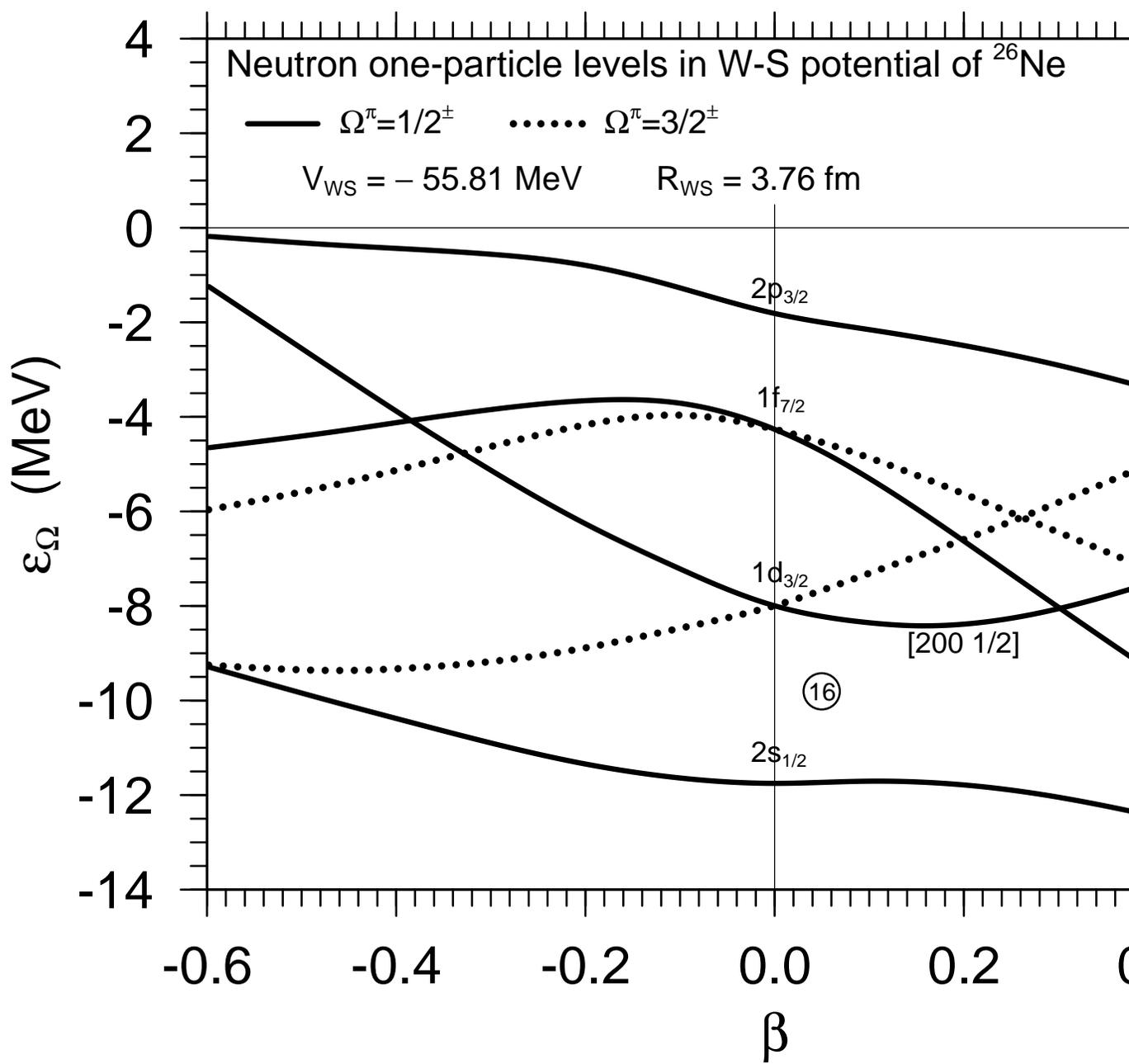
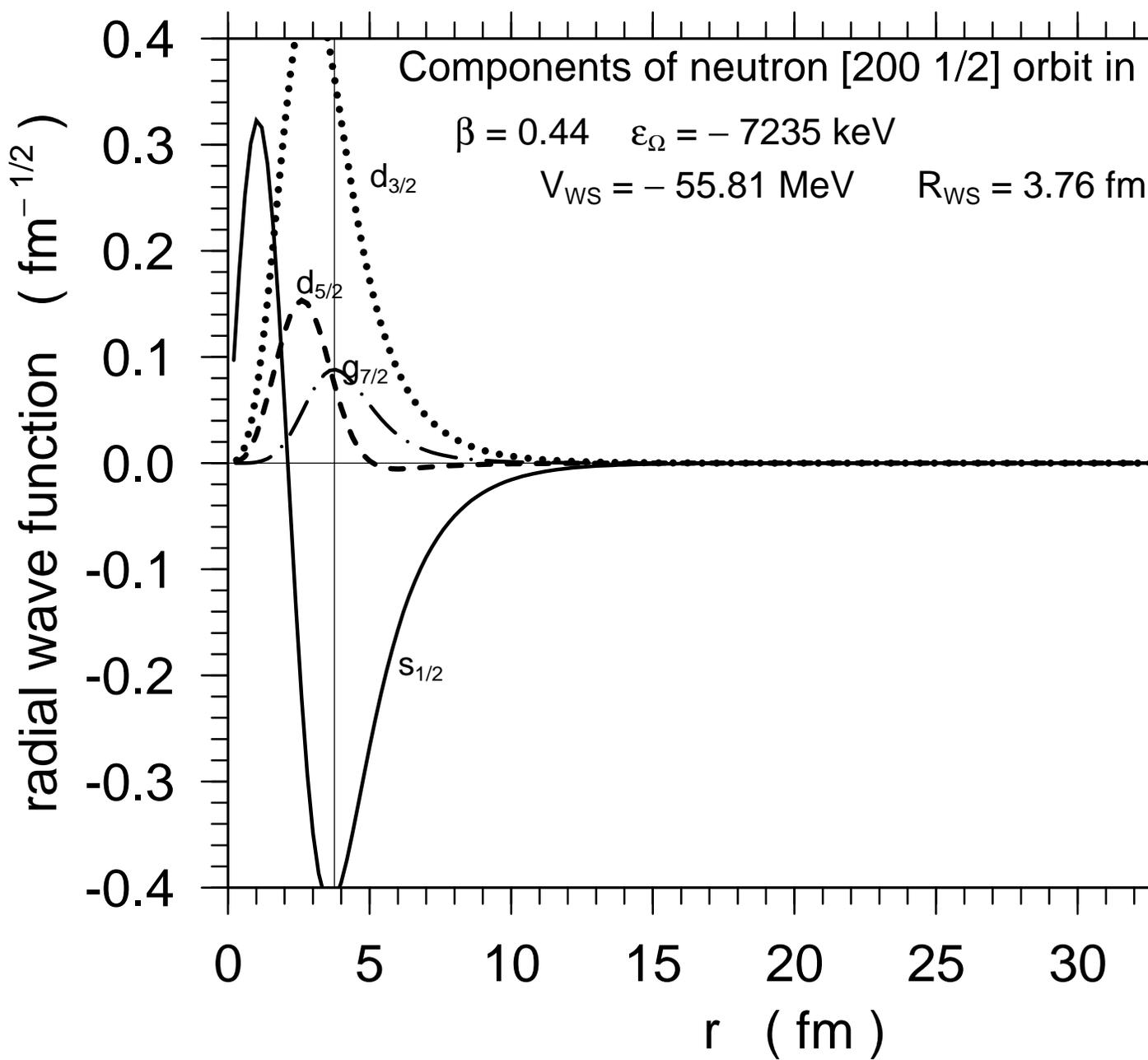
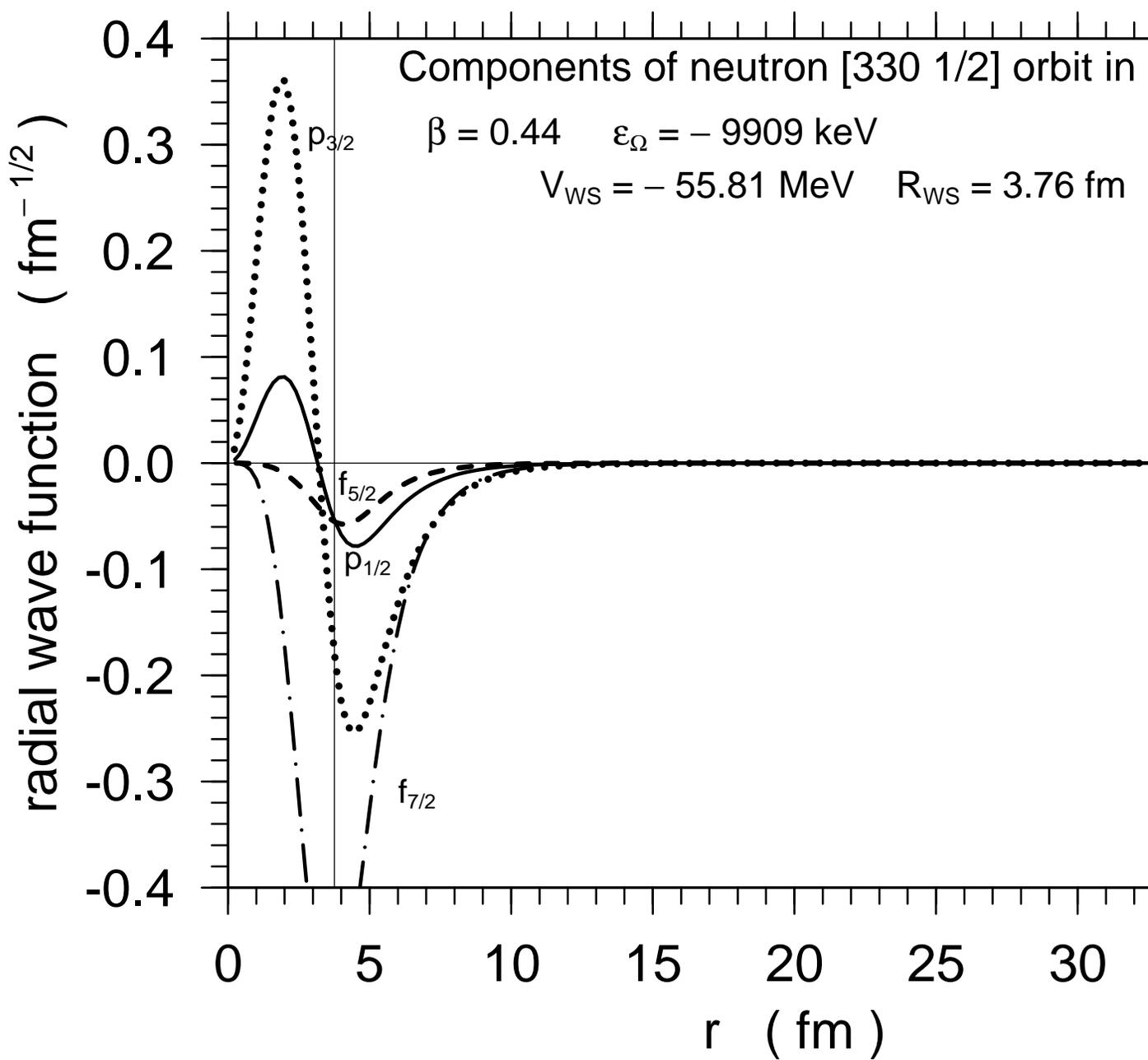
\end{document}